\documentclass[%
 reprint,
superscriptaddress,
nobibnotes,
 amsmath,amssymb,
 aps,
prl,
longbibliography,
citeautoscript,
floatfix,
showkeys
]{revtex4-2}

\usepackage[T1]{fontenc} 
\usepackage[]{graphicx}
\usepackage[utf8]{inputenc}
\usepackage{blindtext}
\usepackage{lmodern}%
\usepackage{hyperref}%
\usepackage{xcolor}%
\usepackage{placeins}
\usepackage{amsmath,amssymb}
\usepackage{bm}
\usepackage{tabularx}
\usepackage{booktabs}
\usepackage{notes2bib}
\usepackage[normalem]{ulem}
\usepackage[normalem]{ulem}
\usepackage{placeins}
\usepackage{amssymb}

\newcommand{\beq}{\begin{equation}\begin{aligned}}
\newcommand{\eeq}{\end{aligned}\end{equation}}

\newcommand{\figref}[2]{\ref{#1}\textsf{#2}}

\newcommand{\ie}{\emph{i.e.},}

\newcommand{\dqmp}{Department of Quantum Matter Physics, University of Geneva, 24 Quai Ernest Ansermet, CH-1211 Geneva, Switzerland}
\newcommand{\gap}{Department of Applied Physics, University of Geneva, 24 Quai Ernest Ansermet, CH-1211 Geneva, Switzerland}
\newcommand{\equal}{These authors contributed equally to this work}

\definecolor{linkcol}{rgb}{0,0,0.4}
\definecolor{citecol}{rgb}{0.5,0,0}
\definecolor{harvardcrimson}{rgb}{0.79, 0.0, 0.09}
\definecolor{lava}{rgb}{0.81, 0.06, 0.13}

\hypersetup{
	bookmarksopen=true,
	pdftitle="Positive  oscillating magnetoresistance in a van der Waals antiferromagnetic semiconductor",
	pdfauthor="Xiaohanwen Lin et al",
	pdfsubject="", 
	pdfstartview={FitH}, 
	pdfhighlight=/O, 
	colorlinks=true, 
	pdfpagemode=UseNone, 
	pdfpagelayout=SinglePage, 
	pdffitwindow=true, 
	linkcolor=harvardcrimson, 
	citecolor=harvardcrimson, 
	urlcolor=lava
}

	\begin{abstract} 
In all van der Waals layered antiferromagnetic semiconductors investigated so far a negative magnetoresistance has been observed in vertical transport measurements, with characteristic trends that do not depend on applied bias. Here we report vertical transport measurements  on layered antiferromagnetic semiconductor  CrPS$_4$ that exhibit a drastically different behavior, namely a strongly bias dependent, positive magnetoresistance that is accompanied by pronounced oscillations  for devices whose thickness is smaller than 10~nm. We establish that this unexpected behavior originates from  transport being  space-charge limited, and not injection limited as for layered antiferromagetic semiconductors studid earlier. Our analysis indicates that the positive magnetoresistance and the oscillations only occur when electrons are injected into in-gap defect states, whereas when electrons are injected into the conduction band the magnetoresistance vanishes. We propose a microscopic explanation for the observed phenomena that combines concepts typical of transport through disordered semiconductors with known properties of the CrPS$_4$ magnetic state, which captures all basic experimental observations. Our results illustrate the need to understand in detail the  nature of transport through vdW magnets, to extract information about the nature of the order magnetic states and its microscopic properties.
 \end{abstract}

\begin{document}
	

	\title{Positive  oscillating magnetoresistance in a\\ van der Waals antiferromagnetic semiconductor}

 \author{Xiaohanwen Lin} 
 \email{Xiaohanwen.Lin@unige.ch}
 \homepage{\equal}
  \author{Fan Wu}
  \homepage{\equal}
   \affiliation{\dqmp}
 \affiliation{\gap}
\author{Nicolas~Ubrig}
\author{Menghan Liao}
\author{Fengrui Yao}
\author{Ignacio Gutiérrez-Lezama} %
\author{Alberto F. Morpurgo}
\email{alberto.morpurgo@unige.ch}
 \affiliation{\dqmp}
 \affiliation{\gap}
 	\date{\today}

 \maketitle

\section*{Introduction}
In most magnetic systems, the application of an external magnetic field causes a better spin alignment and facilitates electron conduction~\cite{spintronic,Nobel}. As a result, the resistance  decreases upon the application of a magnetic field~\cite{PhysRevLNobel,PhysRevLMetalspin,ScienceClosal,PhyLAlo,Rareearth,Naturespin,oxidemagneticmaterial,tunnebarrier1,tunnebarrier2,SpinAlbert}, which is the reason  why giant~\cite{PhysRevLNobel} and colossal magnetoresistive systems~\cite{ScienceClosal},  ferromagnets near the Curie temperature~\cite{Naturespin}, or magnetic tunnel barriers commonly exhibit a negative magnetoresistance~\cite{PhyLAlo,tunnebarrier1,tunnebarrier2}. Indeed, in magnetic conductors, a positive magnetoresistance originating from the coupling of the applied magnetic field to the magnetic state of the materials is only rarely observed~\cite{pMR-Mn,longgen,manyala_addendum_2000}, and when observed the effect is often relatively small~\cite{longgen,manyala_addendum_2000}. These considerations hold true also  for the vast majority of van der Waals (vdW) magnetic materials that  are attracting considerable attention, in which magnetotransport has been investigated either in field-effect transistors or --more commonly-- in tunnel barrier devices (\ie\ using multilayers of magnetic semiconductors as tunnel barriers)~\cite{wang_very_2018,klein_probing_2018,song_giant_2018,kim_one_2018,wang_electric-field_2018,klein_enhancement_2019,song_switching_2019,wang_determining_2019,long_persistence_2020,telford_layered_2020,wu_quasi-1d_2022,wu_magnetotransport_2022,yao2023multiple}.

Many of the vdW magnets studied to date are layered antiferromagnetic semiconductors, \ie\ materials whose layers are uniformly magnetized with the direction of the magnetization alternating from one layer to the next. Examples  of semiconducting layered antiferromagnets include CrI$_3$ multilayers~\cite{huang_layer-dependent_2017,song_giant_2018}, CrCl$_3$~\cite{wang_determining_2019}, two different allotropes of CrBr$_3$~\cite{wang_magnetization_2021}, and CrSBr~\cite{CSB2020,ye2022layer,PhysRevResearch.X.Lin}. The vertical transport properties of all of these compounds show distinctive, common characteristics trends. The resistance always decreases upon increasing the applied magnetic field below the spin-flip field, \ie\ the magnetoresistance is indeed negative. The decrease of the resistance is monotonous,  and occurs either through sharp jumps  or smoothly, depending on whether magnetic anisotropy is strong or weak. The evolution of the magnetoresistance with magnetic field and temperature is the same irrespective of the voltage applied --as changing the bias only affects the absolute magnitude of the magnetoresistance-- and allows mapping the magnetic phase diagram of these systems~\cite{wang_very_2018,wang_determining_2019,soler-delgado_probing_2022}.

Here we report vertical transport measurements on multilayers of layered antiferromagnetic semiconductor  CrPS$_4$ that exhibit an unexpected magnetoresistance at odds with the observations that have been reported  in all other vdW layered antiferromagnets  studied earlier. The evolution of the resistance with magnetic field depends strongly on the applied bias. For small and intermediate bias voltage, the magnetoresistance is large and positive for magnetic field below the spin-flip field, and it becomes negative at larger field. In this bias range, the positive magnetoresistance is accompanied by pronounced magnetoresistance oscillations in devices with thickness below approximately 10~nm. Upon increasing the applied bias, the amplitude of the magnetoresistance and of the oscillations  decreases and eventually vanishes. We show that the unexpected observed behavior is rooted in the fact that vertical transport in  CrPS$_4$ occurs in the space-charge limited regime, \ie\ transport is not injection limited (see top panel of Fig.~\figref{fig1}(a)) as in previously studied layered antiferromagnets. We further show that the different behavior of the magnetoresistance as a function of bias depends on whether space-charge limited transport occurs in the trap-limited (see middle panel of Fig.~\figref{fig1}(a)) or in the trap-free regime (see bottom panel of Fig.~\figref{fig1}(a)). These conclusions allow us to propose a realistic scenario to explain the unexpected phenomena observed in the experiments in terms of defect-mediated transport processes well-established for disordered semiconductors, combined with properties of the magnetic state of  CrPS$_4$ multilayers.

\section*{RESULTS and ANALYSIS}
CrPS$_4$ is a weakly anisotropic layered antiferromagnet, whose easy axis points perpendicular to the layers~\cite{louisy_physical_1978,pei_spin_2016,zhuang_density_2016,peng_magnetic_2020,gu_photoluminescent_2020,calder_magnetic_2020,deng_two-dimensional_2021}. The material becomes antiferromagnetic below T$_N$ = 38~K, and exhibits low-temperature spin-flop and spin-flip transitions at approximately 0.8~T and 8~T~\cite{peng_magnetic_2020,wu_gate-controlled_2023,wu_magnetotransport_2022,wu2023magnetism}. The precise values depend on the batch of crystals employed (CrPS$_4$ crystals were purchased from HQ Graphene), as we found that different batches exhibit different doping levels (inferred from the different values of the linear conductivity measured at high temperature). High-quality field-effect transistors based on multilayers have been already reported, and have allowed the study of magnetism in the material—as well as its dependence on accumulated charge carrier density– by means of in-plane transport~\cite{wu_magnetotransport_2022,wu2023magnetism}. Vertical transport --i.e., transport with current that flows in the direction perpendicular to the layers--has been so far studied only through tetralayer CrPS$_4$, in which direct tunneling gives the dominant contribution to the electrical conductance~\cite{huang2023layer}. The devices that we investigate rely on thicker CrPS$_4$ multilayers (thickness ranging from 4 to 100~nm, with negligible direct tunneling).\\

\begin{figure}[ht]
  \includegraphics[width=1\linewidth]{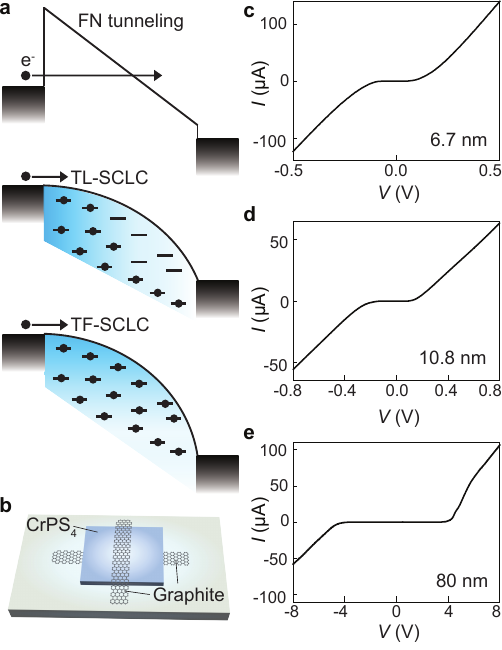}
  \caption{(a) Cartoon schematics of the band diagrams representative of Fowler-Nordheim tunneling (top panel) and space charge limited current  (middle and bottom panel) transport. When Fowler-Nordheim tunneling dominates, the applied voltage generates an electric field that tilts the conduction band and increases tunneling probability. The resulting $I-V$ curves exhibit a very strong exponential dependence of current on bias. In the absence of a sizable barrier at the contacts, transport occurs in the space-charge limited current (SCLC) regime, in which the applied bias both inject electrons in the material and accelerates them. SCLC transport an either be trap-limited (middle panel), when the injected electrons propagate through in-gap defect states (traps), or trap-free (bottom panel), when the traps are filled and the largest contribution to the current is due to electrons  injected into the conduction band. (b) Schematic representation of our graphite/CrPS$_4$/graphite vertical junctions, which are encapsulated between top and bottom hexagonal-boron nitrate (h-BN) layers (not shown); (c)-(e) Current-voltage ($I-V$) characteristics of a 6.7 (c), 10.8 (d) and 80~nm (e) thick device measured at 2, 4 and 4~K respectively.}
  \label{fig1}
\end{figure}

The structures (see Fig.~\figref{fig1}(b) for a schematic representation) were assembled in a glove box with sub ppm concentration of water and oxygen, and encapsulated between hBN layers. The device realization relies on by now standard layer exfoliation and manipulation techniques, and consist of a CrPS$_4$ multilayer in between two multilayer graphene strips attached to the opposite surfaces and acting as contacts~\cite{wang_one-dimensional_2013}. The graphene strips are contacted by metal electrodes which are used to interface the structures with the voltage sources and current amplifier employed to measure their $I-V$ characteristics. More details about the details of device fabrication are presented in the Section S1 of the Supplementary Information. A total of 10 devices have been investigated and show a consistent behavior that we illustrate here by showing data measured on a number of different selected devices. \\

\begin{figure}[htbp]
  \includegraphics[width=1\linewidth]{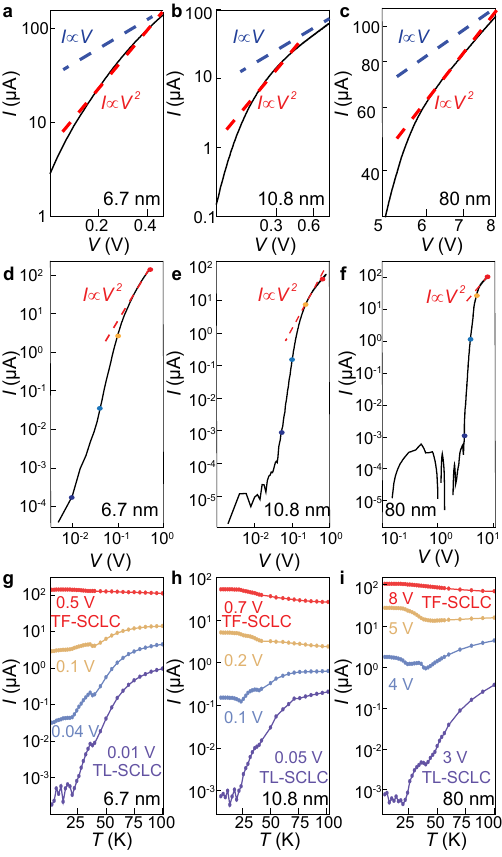}
  \caption{Panels (a)-(f): double-logarithmic plots of the $I-V$ characteristics shown in Fig .~\figref{fig1} (c)-(e) ((a) and (d) 6.7 nm device; (b) and (e) 10.8 nm device; (c) and (f) 80 nm device). Panels (a-c) are zoom-ins of the high voltage bias parts of the $I-V$ curves. The red (slope 2) and blue (slope 1) are guides for the eyes. In panels (a) and (c) the red dashed line overlaps with the data at high bias (i.e., $I \propto V^2)$. In panel (b) the slope at high bias is smaller than 2 and seems to approach one as $V$ is increased, due to the series resistance of the long graphene strip used as contact (see discussion in the main text, and Supplementary Fig. S4 for a comparison of two- and four-terminal measurements on another device, illustrating the effects of the contact resistance).  Panels (g)-(i): temperature dependence of the current measured at different biases for the three devices whose $I-V$ curves are shown in panels (d)-(f) (see legend for device thickness; purple, blue, yellow and red curves represent measurements taken at biases corresponding to the dots of the same color in panels (d)-(f)) . At low bias, the devices exhibit an insulating behaviour (the current decreases with decreasing $T$), whereas at sufficiently high bias the behavior is metallic  (the increases with increasing $T$). } 
  \label{Fig2}
\end{figure}

Fig.~\figref{fig1}(c), (d) and (e) show the current-voltage (I-V) characteristics of three devices (thickness of the CrPS$_4$ layer ranging from 6.7~nm to 80~nm) measured at low temperature ($T=2$ K in panel (c);  $T=4$ K in panels (d) and (e)). The $I-V$ curves exhibit a clear non-linearity, which is nevertheless much less pronounced as compared to most layered antiferromagnetic semiconductors studied earlier~\cite{song_giant_2018,wang_electric-field_2018,kim_tailored_2019}, in which the current increases exponentially fast with bias. To appreciate the difference, we analyze the $I-V$ characteristics by plotting them in a double logarithmic scale (see Fig.~\figref{Fig2}(a-f) and Supplementary Fig. S3). The double- logarithmic plot puts in evidence  the presence of multiple regimes with the current depending differently on voltage  (Fig.~\figref{Fig2}(d-f) and Supplementary Fig. S3). In the lowest voltage range --i.e., between 1 and 10 mV-- $I$  is either linear in $V$ or too small to be measured (i.e., for thicker multilayers the current is below the sensitivity of our amplifiers). For larger voltages, $I$ increase very rapidly with increasing $V$, compatibly with  a power law $I \propto V^m$ with a very large power ($m \approx 10$).  Eventually the double-logarithmic plot tends to flatten resulting in a  $I \propto V^2$ dependence (see the zoomed-in plots in Fig.~\figref{Fig2}(a) and (c), and Supplementary Figure S3 for data from additional devices showing that $I \propto V^2$ at high bias).\\

In a minority of cases the power of the $I \propto V^m$ relation decrease below 2 at the highest bias, and the $I-V$ relation tends to approach linearity (as visible, for instance, in Fig.~\figref{Fig2}(b)). This  happens when the series resistance of the graphene strip used as electrode --larger than 10 k$\Omega$ for the longest contact strips employed in the device fabrication-- cannot be neglected. To illustrate the role of the series resistance we have looked at one of the devices that we investigated, in which  multiterminal measurements could be done to compare the resistance value measured with and without the contribution of the resistance of the graphene strip (in a four-terminal configuration we  measure only the voltage drop across the CrPS$_4$ multilayer). Data from this device are shown in Supplementary Fig. S4. It is clear that at high bias the the slope of the  $I-V$ curve decreases below 2  for measurements done in a two-terminal configuration (i.e., when the resistance includes the series resistance of the graphene strip used as electrode) but the curve  is perfectly quadratic when measured in a four-terminal configuration (to exclude the series resistance and probe exclusively vertical transport through the CrPS$_4$ multilayer).\\

Finding that at high bias the $I-V$ characteristic of the CrPS$_4$ multilayers becomes quadratic provides a first, clear indication that at low temperature vertical transport through CrPS$_4$ multilayers occurs in the so-called space-charge limited current (SCLC)~\cite{PhysRevSCLC,SCLC} regime. This is a different transport mechanism from that observed in similar structures realized to probe vertical transport in other layered antiferromagnetic semiconductor studied earlier. When transport is mediated by SCLC, the role of the applied bias is twofold: it injects charge carriers into the material (just like when charging a capacitor) and it accelerates them. Different regimes of space-charge limited transport are realized depending on whether the charge is injected into in-gap localized states associated to defects (traps), or into states inside a band~\cite{PhysRevSCLC,SCLC}. In the first case –-referred to as trap-limited SCLC–-  the current decreases with lowering temperature and eventually saturates at low $T$, when electron motion is mediated by elastic hopping processes. In the second case –-referred to as trap-free SCLC–- the current  increases upon cooling, since for electrons occupying states in a band the mobility is normally larger at lower $T$. That is why, to confirm that what we observe in our measurements is effectively space charge limited transport, we look at the evolution of the current with temperature at different values of the applied bias $V$.\\

Fig.~\figref{Fig2}(g), (h) and (i) show the temperature dependence of the current measured  for four values of applied voltage, spanning across the different regimes observed in the double-logarithmic plot of the $I-V$ characteristics. It is apparent that at low bias the current $I$ decreases rapidly upon lowering $T$ --a  behavior typical of thermally activated or variable range hopping. As the applied bias is increased, the current still decreases upon cooling, but eventually tends to saturate at low temperature to a value that is sufficiently large to be detected. Such a temperature independent transport regime is expected when transport is mediated by elastic hopping. At sufficiently large bias the current increases upon lowering temperature. Such a behavior, indicative of band transport, starts to occur as the applied bias is such that the $I-V$ curve in the double logarithmic plot enters the $I \propto V^2$ regime. Microscopically, band transport occurs when the non-equilibrium electron distribution generated by the applied bias, and commonly described by a so-called quasi Fermi level, results in a sizable population of states in the conduction band. Overall, therefore, the  evolution of the current with temperature that we observe experimentally is precisely the one expected for SCLC, passing from the trap-limited transport regime (in the bias range where the current decreases upon decreasing $T$) to the trap-free regime at high bias (where electrons injected in the conduction band dominate the current flow). It is therefore the concomitant observation of a high-bias $I \propto V^2$ regime, and of the correlation between bias and temperature dependence of the current, that allow us  to conclude that vertical transport  occurs in the SCLC regime, irrespective of the multilayer thickness (at least in the thickness range investigated). \\

\begin{figure}[htbp]
  \includegraphics[width=1\linewidth]{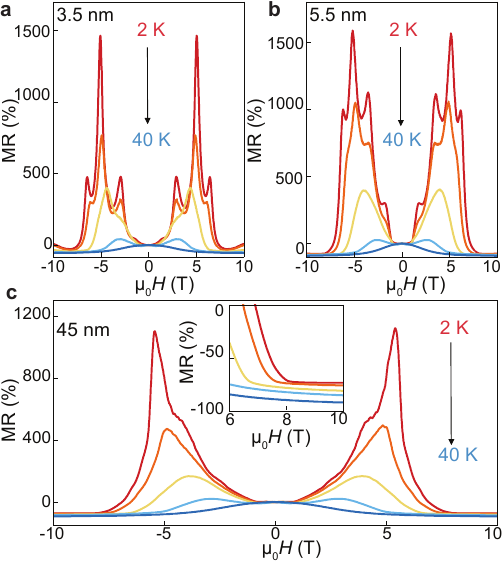}
  \caption{Panels (a)-(c): temperature dependence of the magnetoresistance MR$=\frac{R(H)-R(0)}{R(0)}$ measured at low bias (in the TL SCLC regime) on three different devices with a CrPS$_4$ layer 3.5 (a), 5.5 (b) and 45~nm (c) thick (the magnetic field is applied perpendicular to the layers). Red, orange, yellow, light and dark blue solid lines represent magnetoresistance measured at $T=$ 2, 10, 20, 30 and 40 K, respectively. In all our devices a large (up to between 1000-2000$\%$ at the lowest temperatures) positive magnetoresistance is observed below the spin-flip field ($\simeq$ 8 T), and it exhibits pronounced oscillations in devices that are less than 10~nm thick. The magnetoresstance  becomes negative  and smaller ($\approx100\%$, see inset of (c)), when the field applied is larger than the spin-flip field.} 
  \label{Fig3}
\end{figure}

The behavior of the magnetoresistance of van der Waals magnetic semiconductors operating in the SCLC regime has never been reported earlier, and is currently unknown. We find that at small and intermediate bias the resistance exhibits a pronounced increase corresponding to a positive magnetoresistance between 1000$\%$ and 2000$\%$, before eventually decreasing as $H$ approaches the spin-flip field $\mu _{0} H_{flip}\simeq 8$ T (see Fig.~\figref{Fig3}(a) (b) and (c)). For $H>H_{flip}$, the resistance is lower than at $\mu _{0}H=0$ T, \ie\ the high-field magnetoresistance is negative (with a typical magnitude of $\approx 100 \%$). In devices with thickness smaller than approximately 10~nm, the positive magnetoresistance observed for $H<H_{flip}$ exhibits pronounced oscillations. In thick devices these oscillations are either absent or cannot be clearly resolved and only result  in less pronounced features (e.g., magnetoresistance "kinks"). Such a pronounced  magnetoresistance oscillationss have never been observed earlier in any vdW magnetic semiconductor. \\

\begin{figure}[h]
  \includegraphics[width=1\linewidth]{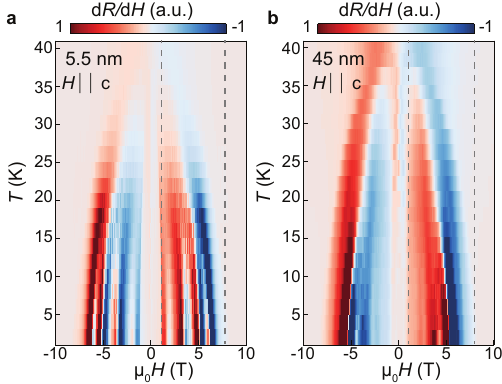}
  \caption{Panels (a)-(b): colour plots of $dR/dH$ calculated from  the conductance $R(H,T)$ measured on the 5.5 (a) and 45~nm (b) thick devices whose  magnetoresistance data are shown in Fig.~\figref{Fig3}(b) and (c), respectively. In all devices, the $dR/dH$ plot show clearly  the spin-flop and -flip fields ($H_{flop}$ and $H_{flip}$, marked by red and black arrow), which shift towards lower values as $T$ increases up to the critical temperature  $T_N\sim 38$ K. The maxima and minima  of the oscillations seen in sufficiently thin devices (panel  (a)) appear as fringes in the color plot, which also move towards lower field values as $T$ is increased, (\ie\ they evolve with $T$ as  $H_{flip}$ does). The dashed vertical lines in the panels represent the spin-flop ($\simeq 0.8 T)$ and the spin-flip ($\simeq 8$ T) transitions observed at the lowest temperature.} 
  \label{Fig4}
\end{figure}

\begin{figure}[ht]
  \includegraphics[width=1\linewidth]{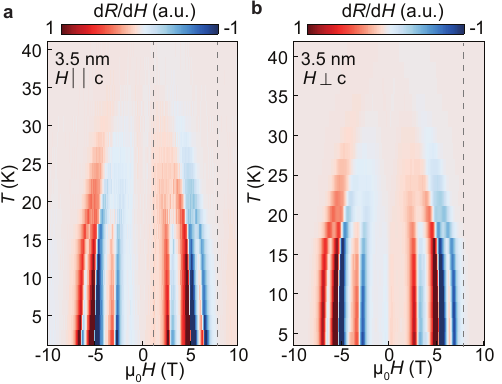}
  \caption{Colour plot of $dR/dH$ as a function of temperature $T$ and magnetic field $H$ measured on a device realized on a 3.5 nm thick CrPS$_4$ multilayer. The data in panels  (a) and  (b) have been measured with the field respectively parallel or perpendicular to the $c$ axis. The dashed vertical lines in the panels represent the spin-flop ($\simeq 0.8$ T) and the spin-flip ($\simeq 8$ T) transitions observed at the lowest temperature.}
  \label{Fig5}
\end{figure}

Upon increasing temperature, the magnetoresistance decreases  and for $T>T_{N}$ --in the paramagnetic state-- only a weak and broad negative magnetoresistance persists (see Fig.~\figref{Fig3}(a), (b) and (c)). The overall evolution as a function of temperature and magnetic field is illustrated in detail in Fig.~\figref{Fig4}~with color plots of  $dR/dH$($H, T$). The data clearly show how $H_{flop}$ and $H_{flip}$ shift to lower values at higher $T$, and eventually vanish at the Néel temperature $T_N\sim 38$ K, as expected. Interestingly, the data further show that magnetoresistance oscillations also evolve with increasing temperature as the spin-flip field does. This is unexpected, because it is well established that in bulk CrPS$_4$ crystals no magnetic transition occurs at the corresponding magnetic field values~\cite{peng_magnetic_2020,wu_magnetotransport_2022}. \\

We find that the overall behavior of the magnetoresistance is nearly identical when the magnetic field is applied perpendicular or parallel to the plane (see Fig.~\figref{Fig5}(a) and (b), as well as Supplementary Fig. S5). This is the case for the complete magnetic field dependence and for its temperature evolution. Indeed, virtually all features observed in the magnetoresistance are present irrespective of whether the field is applied in plane or perpendicular to the plane (with the excpetion of  features originating from the spin-flop transition, which are seen on when the  field is applied perpendicular to the layers, as expected). These observations indicate that  the magnetic field strongly influences the electron motion by coupling to the magnetic spin structure.    \\

\begin{figure}[t]
  \includegraphics[width=1\linewidth]{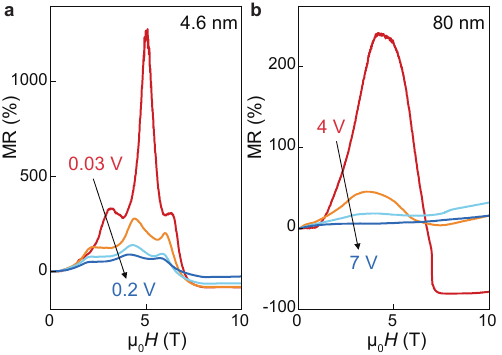}
  \caption{Panels (a) and (b): bias dependence of the magnetoresistance measured in a 4.6 (a) and a 80~nm (b) thick device. The red, orange, light and dark blue solid lines represent the magnetoresistance measured at 0.03 (4), 0.06 (5), 0.12 (6) and 0.2 (7) V in (a) ((b)). The  magnetoresistance is large at low bias, and decreases with increasing bias, eventually vanishing when entering the  TF-SCLC regime (a small magnetoresistance persists in some devices either because the applied bias is not enough to go deeply in the TF-SCLC regime, or because a small contribution from the magnetoresistance of the graphene contacts becomes visible at sufficiently high bias). } 
  \label{Fig6}
\end{figure}

The magnetoresistance also decreases rapidly upon increasing bias. This is shown in Fig.~\figref{Fig6} ~for two devices based on one of the thinnest (Fig.~\figref{Fig6}(a)) and one of the thickest (Fig.~\figref{Fig6}(b)) CrPS$_4$ multilayers studied. It can be seen by comparing the data in Fig.~\figref{Fig6} (b) with the $I-V$ curves for the same devices shown in Fig.~\figref{Fig2} (c) that the bias at which the magnetoresistance disappears corresponds to the onset of the high-voltage range over which $I\propto V^{2}$ (a similar observation has been made on other devices as well). This implies that in the trap-free limit of the SCLC regime, when current is carried by electrons injected in the conduction band of CrPS$_4$, virtually no magnetoresistance is observed, neither positive for $H<H_{flip}$ nor negative negative for $H>H_{flip}$. Indeed, in all devices investigated in which a bias sufficient to enter the trap-free limit was applied, no sizable magnetoresistance was found to survive. At the highest bias, when the resistance is in the 10 k$\Omega$ range, a small positive magnetoresistance due to the contribution of  graphene  is visible in some devices, consistently with the fact --mentioned earlier-- that in some devices at the highest bias transport is limited by the resistance of the graphene strip acting as contact. Finding that the magnetoresistance tends to vanish when the applied bias $V$ bring the device into the $I \propto V^2$ regime allows us to  conclude that the magnetoresistance  (as well as the magnetoresistance oscillations) originate from electrons propagating via localized defect states inside the gap of CrPS$_4$.

\begin{figure}[t]
  \includegraphics[width=1\linewidth]{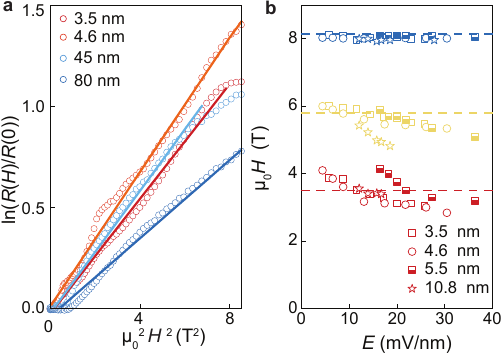}
  \caption{Panel (a): plot of $\ln (R(H)/R(H=0))$ as a function of $H^2$ (with field applied  perpendicular to the layers) measured in several different devices (with the CrPS$_4$ layer respectively 3.5, 4.6, 45 and 80~nm thick). The linear dependence indicates that the positive magnetoresistance measured in the TL-SCLC regime is due to squeezing of the wavefunctions of the states involved in the hopping process. Panel (b): magnetic field values of the two local minima associated to the  magnetoresistance oscillations observed in devices thinner than 10~nm (data taken at 2 K or 4 K, depending on the device; the thickness of the CrPS$_4$ layer is indicated in the legend). The red, yellow and blue data points correspond to the first minimum in the oscillations, the second minimum and the spin-flip field. The dashed lines represent the value of the spin-flip field extracted from CrPS$_4$ transistor measurements done on CrPS$_4$  bilayers (red dashed line), trilayer (yellow dashed line) and thick, bulk-like mulilayers (blue dashed line)~\cite{In}.} 
  \label{Fig7}
\end{figure}

Having established that the positive magnetoresistance originates from electrons hopping between localized states is important, because in conventional (\ie\ non-magnetic) semiconductors a positive magnetoresistance in the hopping regime can be readily explained in terms of wave-function squeezing induced by the applied magnetic field. That is because the orbital effect of the magnetic field on states localized at defects enhances confinement and decreases the  spatial extension of the wavefunctions. As a result, the overlap of the wavefunctions located at neighboring hopping sites decreases, and so does the corresponding hopping probability~\cite{shklovskii2013electronic}. Since at low temperature, the distance of the impurities hosting states with the same energy (\ie\ the states that mediate elastic hopping transport) is large, the wavefunction overlap is due to the  wavefunction tail, whose amplitude decreases exponentially with applied magnetic field. In this regime, theory predict a positive magnetoresistance given by 
$R(H)/R(H=0) = e^{\alpha H^2}$~\cite{shklovskii2013electronic}, such that 
\begin{equation}
    ln( R(H)/R(H=0) ) = \alpha H^2.
    \label{Hopping}
\end{equation}
Indeed if we plot the logarithm of $R(H)/R(H=0)$  versus $H^2$ for devices of based on CrPS$_4$ layers of thickness between 3 and 80~nm, we always  find a linear relation for values of $\mu_{0}H$ up to approximately 3~Tesla (Fig.~\figref{Fig7}(a)).

For non magnetic semiconductors, the magnetoresistance at sufficiently high magnetic field eventually saturates, but for a layered antiferromagnet as CrPS$_4$ the situation is different. That is because in a layered antiferromagnetic semiconductor  increasing the applied magnetic field does not only decrease the extension of the localized states, but also causes  the magnetization of adjacent layers to cant in the same direction. This is important, because when the magnetization in adjacent layers point in opposite directions (for $\mu _{0}H=0$~T) electrons cannot hop from one layer to the next because the exchange energy is too large (approximately 0.6~eV in CrPS$_4$~\cite{wu2023magnetism}). Electrons then need to hop to defect states located in the next-next layer, so that the overlap of the initial and final states involved in the hopping process is truly due to the tail of the wavefunction, and the mechanism for positive magnetoresistance discussed here above is pertinent.\\

As the magnetic field is increased and the magnetization in different layers cant towards the field direction, however, defect states in adjacent layers have a larger parallel spin component. The parallel spin component increases the probability for electrons to hop  from states in one layer to states in the neighbouring layer. Increasing spin-canting, therefore, effectively reduces the distance between states involved in the electron hopping process, and eventually causes the magnetoresistance to start deceasing at sufficiently large  magnetic field. That is why the observed magnetoresistance in CrPS$_4$ devices  peaks  below the spin-flip field. Eventually, for $H$ above the spin-flip field, the magnetization of all layers points in the same direction, and hopping is fully dominated by states in the neighbouring layers, which have a much larger overlap. The closer distance between the defect states means that  the wavefunction overlap is  not  determined by their exponential tails, and is therefore not anymore exponentially sensitive to an increase in magnetic field. In this regime, magnetotransport is likely dominated by a different process, namely the shift to lower energy of the conduction band edge that has been recently observed in field-effect transistor measurements~\cite{wu_gate-controlled_2023, wu2023magnetism}. The lowering of the conduction band edge  increases the spatial extension of the localized wavefunctions relatively to the $\mu _{0}H=0$~T case, resulting in a negative magnetoresistance for $H>H_{flip}$.

The experimental observations indicate that the role played by  spin-canting in adjacent layers to determine the relevant hopping processes is likely also the reason for the magnetoresistance oscillations observed in sufficiently thin CrPS$_4$ layers. This is suggested by the analysis of the magnetic field at which the resistance exhibit minima in devices thinner than approximately 10~nm. The position of these magnetoresistance minima (as a function of applied electric field) is summarized in Fig.~\figref{Fig7}(b), based on data measured on four different devices with thickness smaller than 10~nm. We focus on the minima that are clearly defined, so that their position can be determined unambiguously (more shoulders or kinks become visible as the CrPS$_4$ thickness increases, but the positions of those features are more difficult to determined precisely). In all devices we find two such minima, at $\mu _{0}H\simeq 3.5-4$~T and $\mu _{0}H \simeq 6$~T. We also plot the value of the spin-flip field $\mu _{0}H\simeq 8$~T, at which the magnetoresistance saturates to a constant value.
Interestingly, the values of $\mu _{0}H$ for the two minima identified match quite precisely the spin-flip field of bi- and tri-layer CrPS$_4$, which we know from separate studies of MR on CrPS$_4$ field-effect transistors~\cite{In}. This observation suggests that inside the multilayers used to study vertical transport structural domains are present, with stacking faults that decouple magnetically the CrPS$_4$ layers.

Structural defects (such as stacking faults) can cause electronic decoupling of thin layers within  thicker crystals, resulting  in CrPS$_4$ regions that effectively behave as bi, tri, tetralayers. In terms of their magnetic response, these layers would have a correspondingly lower spin-flip field, as expected theoretically. For bilayers, for instance, the spin-flip field is half that of bulk, because each constituent layer feels the exchange interaction of only one other neighbouring layer, and not of two like in the bulk (for a study of spin-flip field as a function of multilayer thickness in low-anisotropy layered antiferromagnets see past work on CrCl$_3$). We therefore suggest that the observed peaks in the magnetoresistance originate from the contributions of regions inside the device in which stacking faults cause electronic decoupling of few-layers, which contribute  to the total measured resistance. Specifically, because of the lower spin-flip field, in these regions a large spin canting  starts at smaller $H$, causing their contribution to the total  magnetoresistance to also peak at values of  $H$ smaller than that of thick multilayers. To confirm this idea --and explore whether electronic decoupling caused by structural defects may be controlled to  engineer the magnetoresistive response-- it  would certainly be interesting to investigate crystals of different structural quality.\\

\section*{CONCLUSIONS}
In devices based on vdW layered antiferromagnetic semiconductors studied in the past, transport was virtually always injection limited and the experimental results were properly captured in terms of a phenomenological Fowler-Nordheim tunneling  description ~\cite{fowler1928electron,lenzlinger1969fowler}. This same approach, however, does not properly describe the behavior of   CrPS$_4$ based devices,  because vertical transport in CrPS$_4$ occurs in the space charge limited current regime, \ie\ electron injection at the contacts is not the process limiting current flow. Why vertical transport occurs in the SCLC regime in CrPS$_4$ multilayers (whereas for all other 2D magnetic materials studied earlier at high bias transport was found to be injection limited) remains to be understood in detail. It is probably due to the high (unintentional) doping level of CrPS$_4$ crystals,  which shifts the Fermi level in the material much closer to the conduction band edge as compared to other 2D magnetic van der Waals semiconductors. The closer proximity of the Fermi level to the conduction band edge is corroborated by the observation of a linear transport regime in many of our devices down to low temperature, with relatively small  activation energy of the resistance (30 to 40 meV depending on crystals, see Supplementary Fig. S2), much smaller than for instance in CrI$_3$ (where the activation energy is approximately 200 meV)~\cite{wang_very_2018}.  The proximity of the Fermi level to the conduction band edge  enhances both the density of states at the Fermi level and the overlap of the  wavefunctions of electrons localized at defects, drastically increasing conductivity  due to in-gap states through CrPS$_4$. The conductivity due to in-gap state implies that charge injected in the material is sufficiently mobile to enable the observation of current, which is precisely the idea of space charge limited transport. In other vdW magnets, in contrast, charge injected into in-gap states inside the material does not lead to any measurable current, because  the much larger distance between Fermi level and conduction band edge implies that electrons in defects are much more strongly localized. As a consequence, current is only observed when the bias is large enough to inject directly electrons into the band, which is the idea behind Fowler-Nordheim tunneling.

Irrespective of the precise microscopic reason why vertical transport in CrPS$_4$ is space charge limited, what is particularly interesting is that the SCLC transport regime has not been investigated earlier in layered vdW magnets, and nothing is known about the expected behavior of the magnetoresistance. In the space charge limited transport regime, the applied bias accumulates electrons in the material --similarly to what happens in a capacitor-- and it then accelerates them. At low bias, electrons are injected into localized states, and can propagate only via hopping. Upon increasing the bias, the larger density of accumulated electrons start to also occupy states in the conduction band. Electrons occupying these states can propagate much more easily and cause a large decrease in resistance. The microscopic processes responsible for current flow are therefore  different at low and high bias. This difference  is the reason for the strong bias dependence of the magnetoresistive response and of the temperature dependence of the measured resistance. Because CrPS$_4$ devices are the first in the family of vdW semiconducting magnets to operate in the space charge limited transport regime, their behavior is  different from that observed in materials studied earlier, and  appears anomalous at first sight.

Our work succeeds in  rationalizing all the seemingly anomalous experimental observations by considering the interplay between semiconductor physics --which determines how the localized states responsible for hopping respond to an applied magnetic field-- with the evolution of the magnetic state of CrPS$_4$ upon increasing the applied magnetic field --which determines whether electrons can or cannot hop to states in an adjacent layer. The application of magnetic field enhances localization effects of electrons in in-gap states involved in the hopping conduction, causing the resistance to increase at low field. The gradual increase in canting of the  magnetization of different layers in a same direction enhances the probability for electrons to hop from one layer to the nearest neighboring one, which effectively decreases the distance between hopping sites. The net result is the positive and non-monotonic magnetoresistance that we observe. Once the accumulated electrons populate states in the band, the resistance becomes much less sensitive to magnetic field. The observed overall decrease in resistance relative to the value measured at $\mu _{0}H=0$~T is likely again due to the down shift of the conduction band edge.

In summary, our results show how vertical magnetoresistance measurements on van der Waals semiconducting magnets can exhibit a very different evolution with field, temperature and bias in different  materials with virtually identical magnetic states, simply because of the different regimes in which transport occurs. Our experiments also reveal the behavior of the magnetoresistance in the space charge limited transport regime, that in van der Waals semiconducting magnets had never been investigated earlier. Both these issues are important to put on solid grounds magnetoresistance measurements as a reliable technique to investigate the magnetic properties of atomically thin van der Waals magnets. \\

\section*{Acknowledgements}
The authors gratefully acknowledge Alexandre Ferreira for continuous and valuable technical support, and M. Gibertini, J. Fernandez-Rosier and I. Martin for interesting discussions. AFM Acknowledges financial support from the Swiss National Science Foundation Division II under project 200021-219424.\\

\FloatBarrier

%

\end{document}